\documentclass[twocolumn,showpacs,preprintnumbers,amsmath,amssymb]{revtex4}

\usepackage{graphicx}
\usepackage{dcolumn}
\usepackage{bm}

\newcommand {\bk}{{\bf k}}
\newcommand {\cP}{{\cal P}}

\begin{document}


\title{Inelastic electron relaxation rates caused by Spin M/2 Kondo Impurities}

\author{Georg G\"oppert and Hermann Grabert}
 \affiliation{
 Physikalisches Institut, Albert--Ludwigs--Universit{\"a}t,  \\
 Hermann--Herder--Stra{\ss}e~3, D--79104 Freiburg, Germany
 }

\date{13 Mai 2003}

\begin{abstract}
We study a spin S=M/2--Kondo system coupled to electrons in
an arbitrary nonequilibrium situation above Kondo temperature.
Coupling to hot electrons leads to an increased inverse
lifetime of pseudo particles, related to the Korringa width. This
in turn is responsible for the increased inelastic relaxation
rates of the electronic system.
The rates are related to spin--spin
correlation functions which are determined using a projection
operator formalism. The results generalize recent findings
for S=1/2--Kondo impurities which have been used to describe
energy relaxation experiments in disordered mesoscopic wires.
\end{abstract}


\maketitle

\vspace{-0.2truecm}
Recently, experimental evidence was found that Kondo
impurities might play an essential role for energy relaxation
in mesoscopic gold wires \cite{PierreERelaxPecs01} displaying 
much higher energy relaxation rates than predicted
by standard theory \cite{AltshulerEEint85}.
Based on these findings several theoretical studies have led to
a qualitative or even quantitative explanation of experimental 
data by accounting for electron--electron interaction 
mediated by magnetic impurities
\cite{GlazmanERelaxPRL01,GeorgErelaxPRB01,KrohaPecs01,KrohaNEQPRL02}.
Assuming Kondo impurities of unknown origin as relevant inelastic
scattering centers also earlier experimental findings
on copper wires
\cite{PothierDistrPRL97}
could be explained
\cite{GeorgErelaxPRB01,KrohaPecs01,KrohaNEQPRL02}.
Moreover, assuming spin $1/2$--impurities \cite{GeorgErelaxPRB02}
the detailed magnetic field dependence of energy relaxation
experiments on copper wires \cite{PierreErel01} could be
fitted, strongly suggesting that Kondo impurities
indeed play an essential role for energy relaxation
at low temperatures.

In a recent work Anthore {\it et al.}
\cite{AnthoreAgWiresPRL03} studied energy relaxation in
thin silver wires with Mn impurities and explained their findings
using both, direct electron--electron interaction
\cite{AltshulerEEint85} and the effect of spin $1/2$--impurities
\cite{GeorgErelaxPRB02}. Since Mn in silver is not a spin
$1/2$--impurity and furthermore the spin of the impurities
in copper is not known, a generalization of the theory
in Ref.~\cite{GeorgErelaxPRB02} is desirable. 
In addition the impurity densities $c_{\text{imp}}$ gained by
fitting the energy relaxation data of the copper and silver
samples typically exceed those obtained from measurements of the 
dephasing rate by more than an order of magnitude, see
Refs.~\cite{GeorgErelaxPRB02,AnthoreAgWiresPRL03} and articles 
cited therein. Impurity densities as high as those inferred from 
energy relaxation rates would lead to much
higher dephasing rates than those found in experiments.

Considering the theoretical work in
Refs.~\cite{GlazmanERelaxPRL01,GeorgErelaxPRB01},
the impurity density can be lowered by
increasing the spin $S$ because only the product
$S(S+1) c_{\text{imp}}$ enters the prefactor of
the rate.
However, this
result does not take into account the spin
dependence of the renormalized coupling constant.
The aim of the present work is a generalization of the findings in
Ref.~\cite{GeorgErelaxPRB02} to arbitrary spin thereby 
exploring the possibilities of lowering the impurity density by
increasing the spin $S$.

Since this work is an extension of Ref.~\cite{GeorgErelaxPRB02} 
we follow the argumentation therein and, as far as possible, use 
the same notation. In order to make the paper self--contained, some 
of the basic ideas and definitions are, however, repeated. 
Whereas the technical details change, the
main physical arguments remain the same and we refer the reader to
Ref.~\cite{GeorgErelaxPRB02} for further information.


We describe the quasiparticles and the impurity spin
by the free Hamiltonian
\begin{equation}\label{eq:FreeHamiltonian}
    H_0
    =
    \sum_{\bk \sigma} \epsilon^{}_{\bk \sigma}
    C^\dagger_{\bk \sigma} C^{}_{\bk \sigma}
    -
    E_H S^z
\end{equation}
where $C_{\bk\sigma}^\dagger$ and $C_{\bk\sigma}^{}$
create and annihilate an electron in a given
orbital, $\bk$, and spin, $\sigma$, state.
$\epsilon_{\bk\sigma}$ is the energy of this
state. The second term in
Eq.~$(\ref{eq:FreeHamiltonian})$ describes a
spin $M/2$--impurity
with Zeeman splitting $E_H=g \mu_B B$.
The coupling between quasiparticles and impurity spin
is described by the standard Kondo Hamiltonian
\begin{equation}\label{eq:sdinteracren}
 H_I
 =
 J_0 \sum_{\bk \bk'\sigma \sigma'}
 {\bf S} \cdot {\bf s}_{\sigma' \sigma}
 C^\dagger_{\bk' \sigma'} C^{}_{\bk \sigma}
\end{equation}
where $J_0$ is the bare coupling and
${\bf s}_{\sigma' \sigma}$ denotes the vector of 
Pauli matrices. Here, we assume the impurity
density $c_{\text{imp}}$ to be small enough that we need to 
treat coupling to a single impurity only.

To determine the inelastic electron rates we consider
the angularly averaged
collision integral which in linear order in the
density $c_{\text{imp}}$ reads
\cite{RammerRMP86}
\begin{eqnarray}
 I_\sigma(\epsilon)
 =\frac{i}{\hbar}
  \biggl\{
    f_\sigma(\epsilon) \Sigma_\sigma^>(\epsilon)
    +
    [1-f_\sigma(\epsilon)] \Sigma_\sigma^<(\epsilon)
  \biggr\} \,.
\label{eq:collisiongen2}
\end{eqnarray}
Here,
$\Sigma_\sigma^{>/<}(\epsilon)=\Sigma^{>/<}(\bk \sigma,\epsilon)$
where $\epsilon=\epsilon_{\bk\sigma}$ is the electron self--energy
on shell, assumed to be independent of the
angular momentum. $f_\sigma(\epsilon)$ is the
angularly averaged distribution function for electrons
of energy $\epsilon$ and spin $\sigma$. For readability
we suppressed the spatial dependence. Since the self--energy is
proportional to the impurity density, we already replaced
the electron Green's functions by their
unperturbed form and integrated over frequency to get
the classical form of the collision integral.
In contrast to Ref.~\cite{GeorgErelaxPRB02} we do not
use the spin averaged self--energy but generalize the
results to spin dependent distribution functions.

Our task is now to determine the electron self--energy
which in turn leads to the electron scattering rates.
Changing to the interaction picture and representing
the spin degrees of freedom by pseudo--particles
\cite{Abrikosov65} one can use perturbation
theory on the Keldysh contour to generate the
graphs contributing to the electron self--energy.
Since the topological structure of the graphs for
a spin $1/2$--system  and a spin $M/2$--system is the same,
we can directly follow the reasoning in
Ref.~\cite{GeorgErelaxPRB02}.

In lowest order the electron self--energy is given
by a pseudo--Fermion bubble and an electron or hole
line in between. The pseudo--Fermion bubble can be
represented as a spin--spin correlation function which
in frequency space directly determines the rates.
Higher order corrections are separable into terms
adding an additional electron--hole pair and terms
leading to higher order corrections for a single
electron--hole pair. The combinations of the second 
type are usually referred
to as singe particle intermediate state corrections and
can be absorbed by a renormalization of the coupling
constants \cite{SilversteinPR67}. 
For arbitrary spin $S$, we find that the renormalized
vertices $J_{\pm}^z$, $J^\pm$ only depend on electronic
occupation factors and therefore are given by relations
very similar to those derived in 
Ref.~\cite{GeorgErelaxPRB02}. 
For a non--spin--flip processes we have
\begin{eqnarray}
 J_{\pm}^z(\epsilon)/J_0
\!\! &=& \!\!
 \Big\{
   \left|1
   \!-\! (\pi \rho J_0)^2 S(S+1)/4
   \!-\! \rho J_0 g^{}_\mp(\epsilon \mp E_H)\right|^2
\nonumber \\
&& \qquad
   + (\pi \rho J_0)^2 S(S+1)
\Big\}^{-1/2}
\label{eq:Jupdwren}
\end{eqnarray}
and for a spin--flip process
\begin{eqnarray}\label{eq:Jpmren}
 J^\pm(\epsilon)/ J_0 &=&
  \Big\{
    \big|
     1  - (\pi \rho J_0)^2 S(S+1)/4  - \rho J_0 [g^{}_\pm(\epsilon)
\nonumber \\
&& \hspace*{-0.7cm}
                                  +  g^{}_\mp(\epsilon \pm E_H)]/2
    \big|^2
   + (\pi \rho J_0)^2 S(S+1)
  \Big\}^{-1/2} \! . \quad
\end{eqnarray}
The renormalization is
determined by the auxiliary function
\begin{equation}\label{eq:gfunction}
 g^{}_\pm(\epsilon)
 =
 \int_{-D}^{D} d\epsilon'
 \frac{f^{}_\pm(\epsilon')-1/2}{\epsilon-\epsilon'+i\delta}
 \, .
\end{equation}
In equilibrium this leads to the usual logarithmic corrections,
however, the above formulae are applicable for arbitrary 
nonequilibrium situations.
The Kondo temperature in this approximation reads
\begin{equation}\label{eq:KondoTemp}
 T_K
 =
 D\exp\left\{
 -\frac{1}{\rho J_0}\left[1-\frac{(\pi \rho J_0)^2 S(S+1)}{4}\right]
     \right\}
\end{equation}
and equals the bulk Kondo temperature. The phrase ``above Kondo
temperature'' in this work means
that the corrections determined by the auxiliary function
$(\ref{eq:gfunction})$ are still small compared to one.
In this sense a system below the equilibrium Kondo temperature 
can be ``above Kondo temperature'' because of the 
nonequilibrium smearing of the distribution function.

Well above Kondo temperature it is usually assumed that all
vertices renormalize independently. Therefore, one can
equivalently put these renormalized quantities in a new
interaction Hamiltonian
\begin{eqnarray}
 H_I
\!\!&=&\!\!
 \frac{1}{2} \sum_{\bk \bk'}
 \bigg\{
   S^+ J^+(\epsilon_{\bk \uparrow})
   C_{\bk' \downarrow}^\dagger C_{\bk \uparrow}^{} +
   S^- J^-(\epsilon_{\bk \downarrow})
   C_{\bk' \uparrow}^\dagger C_{\bk \downarrow}^{}
\nonumber \\
&&\!\! +
   S^z
   \left[
     J^z_+(\epsilon_{\bk \uparrow})
     C_{\bk' \uparrow}^\dagger C_{\bk \uparrow}^{}  -
     J^z_-(\epsilon_{\bk \downarrow})
     C_{\bk' \downarrow}^\dagger C_{\bk \downarrow}^{}
   \right]
 \bigg\} \,.
\label{eq:sdinteracren1}
\end{eqnarray}
with energy and process dependent coupling constants.
Using this Hamiltonian we have to restrict to
elementary electron--hole pair excitations only. Other,
more complex graphs of the one--particle intermediate
state correction type, are already put into the
renormalization of the coupling constants.
The electron self--energy is now given by the
pseudo--Fermion bubble coupled to arbitrarily many
simple electron--hole pairs with an electron or hole line
in between and can be written as
\begin{eqnarray}\label{eq:rate_vs_selfen}
 \Sigma_\sigma^>(\epsilon)
 =
 -i \sum_{\sigma'}  \int d\epsilon'
 W_{\sigma,\sigma'}(\epsilon,\epsilon')
 [1-f_{\sigma'}(\epsilon')]
\end{eqnarray}
for the larger self--energy where $W_{\sigma,\sigma'}$ denotes the
corresponding rates. The smaller self--energy
$\Sigma_\sigma^<(\epsilon)$ is given by changing the 
variables, $(\epsilon, \sigma)$ to $(\epsilon', \sigma')$ and 
$f \rightarrow 1-f$. Rewriting the
pseudo--Fermion bubble as spin--spin correlation function, the rates
are given by
\begin{eqnarray}
  W_{-,+}(\epsilon,\epsilon')
&=& \frac{c_{\rm imp} \rho}{4 \hbar}
  J^-(\epsilon) J^+(\epsilon')  C_+(\epsilon-\epsilon')
 \label{eq:Ratedwup} \\
  W_{+,-}(\epsilon,\epsilon')
&=& \frac{c_{\rm imp} \rho}{4 \hbar}
  J^+(\epsilon) J^-(\epsilon')  C_-(\epsilon-\epsilon')
 \label{eq:Rateupdw} \\
  W_{+,+}(\epsilon,\epsilon')
&=& \frac{c_{\rm imp} \rho}{4 \hbar}
  J^z_+(\epsilon) J^z_+(\epsilon')  C_z(\epsilon-\epsilon')
 \label{eq:Rateupup} \\
  W_{-,-}(\epsilon,\epsilon')
&=& \frac{c_{\rm imp} \rho}{4 \hbar}
  J^z_-(\epsilon) J^z_-(\epsilon')  C_z(\epsilon-\epsilon')
 \label{eq:Ratedwdw}
\end{eqnarray}
with
\begin{equation}
 C_\pm(t)
 =
 \langle S^\pm(t)S^\mp(0) \rangle\, ,
\quad
 C_z(t)
 =
 \langle S^z(t)S^z(0) \rangle \, .
\label{eq:correlfkt}
\end{equation}
Using $(\ref{eq:rate_vs_selfen})$ the 
collision integral takes the standard form for 
spin dependent scattering
\begin{eqnarray}\label{eq:CollInt}
 I_\sigma(\epsilon)
&=&
 \sum_{\sigma'}  \int d\epsilon'
 \big\{
  f_{\sigma}(\epsilon)[1-f_{\sigma'}(\epsilon')]
        W_{\sigma,\sigma'}(\epsilon,\epsilon')
\nonumber \\
&& \qquad
  -[1- f_{\sigma}(\epsilon)]f_{\sigma'}(\epsilon')
        W_{\sigma',\sigma}(\epsilon',\epsilon)
 \big\}   \,.
\end{eqnarray}
The energies, $\epsilon=\epsilon_{\bk \sigma}$,
measure the kinetic energy and the Zeeman
energy. Usually, when going over into a
continuum description the Zeeman
splitting is put to a band bottom shift. 

As in Ref.~\cite{GeorgErelaxPRB02}
we use a projection operator formalism to determine the
correlation functions in an arbitrary nonequilibrium situation.
Using the projection operators
\begin{eqnarray}
    P^z X
&=& S^z \langle X S^z \rangle/\langle S^z S^z \rangle
    \qquad \mbox{for} \quad C_z \quad \mbox{and}
    \label{eq:ProjOpz}     \\
    P^\pm X
&=&
    S^\pm \langle X S^\mp \rangle/\langle S^\pm S^\mp \rangle
    \qquad \mbox{for} \quad C_\pm
    \label{eq:ProjOppm}
\end{eqnarray}
one can derive a formally exact integro--differential equation
\cite{GrabertPOT82}
\begin{equation}
 \dot{C}_a(t)
 =
 \Phi_a C_a(t)
 -\int_0^t \!\! du\,  \phi_a(t-u) C_a(u)
\end{equation}
with the solution in terms of the Laplace transform
\begin{equation}
 \tilde{C}_a(z)
 =
 \frac{C_a(t=0)}{z-\Phi_a+\tilde{\phi}_a(z)} \,.
\end{equation}
Here $a=z,\pm$ and, as in the $S=1/2$ case, $\Phi_z=\langle
\dot{S}^z S^z \rangle/\langle S^z S^z \rangle=0$ and $\Phi_\pm
 =
 \langle \dot{S}^\pm S^\mp \rangle/\langle S^\pm S^\mp \rangle
 =
 \mp i \tilde{E}_H$, which
leads to the free propagation, where
$\tilde{E}_H$ includes the Knight shift neglected throughout this
work. The averages are to be calculated self--consistently
together with the steady state
electronic distribution functions $f_\sigma$ 
and the occupation probabilities
$\cP^{}_m$ for the impurity spin being in state $m$.

The memory kernel $\phi_a(t)$ for the $C_\pm$ correlation function
reads
\begin{equation}
 \phi_\pm(t)
 =
 \frac{
   \langle \dot{S}^\pm_r(t) \dot{S}^\mp \rangle
      }{
   \langle S^\pm S^\mp \rangle
       }
 +
 \Phi_\pm
  \frac{
   \langle \dot{S}^\pm_r(t) S^\mp \rangle
      }{
   \langle S^\pm S^\mp \rangle
       } \, .
\label{eq:MemoKernelPM}
\end{equation}
Here, the index $r$ in  $S^\pm_r(t)$ indicates
that the dynamics of the spin operator
is reduced by the projection. It is determined by the expression
$\dot{S}^\pm_r(t)=\exp[i\hat{L}(1-P^\pm)t] \dot{S}^\pm$ with the
Liouville operator $\hat L$ acting as
$\hat{L}\hat{X}=[H,\hat{X}]/\hbar$.
The memory kernel for the $C_z$
correlation function is given by Eq.~$(\ref{eq:MemoKernelPM})$
with the replacements $\pm,\mp \rightarrow z$.
We are interested in the regime well above Kondo temperature
and expand the kernel up to second order in
the renormalized coupling $J$. Since the dynamics of the expanded
kernel function is oscillatory, the Fourier transformed correlation
function has always the simple form
\begin{equation}
 C_a(\omega)
 =
 \frac{2 C_a(t=0) \, {\rm Re} \, \phi_a(\omega)}
 {[\omega -i\Phi_a + {\rm Im}\, \phi_a(\omega)]^2
  +[{\rm Re} \, \phi_a(\omega)]^2}
\label{eq:CorrelGen}
\end{equation}
with $a=z,\pm$.
Further, we define ${\rm Re}  \, \phi_a(\omega)
 \equiv
 {\rm Re}\, \{\tilde{\phi}_a(-i\omega+\delta)\}$ and the
imaginary part ${\rm Im} \, \phi_a(\omega)$ follows from the
Kramers--Kronig relation.
When calculating the electronic distributions $f_\sigma$ or 
spin occupation
probabilities $\cP_m^{}$, the imaginary parts 
${\rm Im}\, \phi_a(\omega)$ in the denominators
lead to higher order corrections in $J$ and are neglected.

The damping rates (which were named $\nu_a(\omega)$ with 
$a=z,\pm$ in Ref.~\cite{GeorgErelaxPRB02}) read
\begin{equation}\label{eq:RePhiz}
    {\rm Re}  \, \phi_z(\omega)
    =
    \frac{\pi}{4}
    \sum_\pm
    \left[
      \frac{\langle S^\pm S^\mp \rangle}{\langle S^z S^z \rangle}
      \zeta_\pm(\omega \mp E_H)
    \right]
\end{equation}
for the $C_z$ correlation function and
\begin{equation}\label{eq:RePhipm}
    {\rm Re}  \, \phi_\pm(\omega)
    =
    \frac{\pi}{4}
    \left[
      \zeta_z(\omega \mp E_H)  +
      4 \frac{\langle S^z S^z \rangle}{\langle S^\pm S^\mp \rangle}
      \zeta_\mp(\omega)
    \right]
\end{equation}
for the $C_\pm$ correlation functions. The auxiliary functions
\begin{equation}\label{eq:ZetaZ}
    \zeta_z(\omega)
    =
    \sum_\pm
    \int d\epsilon \rho^2
      J_\pm^z(\epsilon)J_\pm^z(\epsilon+\omega)
      f_\pm(\epsilon)[1-f_\pm(\epsilon+\omega)]
\end{equation}
and
\begin{equation}\label{eq:ZetaPM}
    \zeta_\pm(\omega)
    =
    \int d\epsilon \rho^2
      J^\mp(\epsilon)J^\pm(\epsilon+\omega)
      f_\mp(\epsilon)[1-f_\pm(\epsilon+\omega)]
\end{equation}
describe coupling to electron--hole pairs. In
equilibrium the damping leads directly to the Korringa width 
proportional to the temperature whereas in
nonequilibrium this rate scales with a measure of the
nonequilibrium situation, namely $eU$, leading to an 
increased inverse lifetime independent of the 
measurement temperature. 

The equal time correlation functions read for $S=M/2$
\begin{equation}\label{eq:SzSzExpValue}
    C_z(t=0)
    =
    \langle S^z S^z \rangle
    =
    \sum_{m=-M/2}^{M/2} \cP^{}_m m^2
\end{equation}
and
\begin{equation}\label{eq:SpSmExpValue}
    C_\pm(t=0)
    \!=\!
    \langle S^\pm S^\mp \rangle
    \!=\!
    \sum_{m=-M/2}^{M/2} \cP^{}_m [S(S+1)-m(m\mp 1)]
  \, .
\end{equation}
Independent of the distribution $\cP^{}_m$ the
spin--spin correlation function $C(t) =\langle {\bf S}(t)\cdot
{\bf S} \rangle =[C_+(t) + C_-(t)]/2+C_z(t)$ fulfills the sum rule
$ C(t=0)= \int (d\omega/2\pi) \,  C(\omega) =S(S+1) \, .$

To determine the master equation for the $\cP^{}_m$'s we use
Eq.~$(\ref{eq:SpSmExpValue})$ and write the spin--flip correlation
function as
\begin{equation}\label{eq:CPMsplitting}
    C_\pm(\omega)
    \equiv
    \sum_m \cP^{}_m [S(S+1)-m(m\mp1)] \check{C}_\pm(\omega)
 \,.
\end{equation}
The rate for the transition from state $m$ to $m\pm 1$ then reads
\begin{equation}\label{eq:GammammPM1}
    \Gamma_{m \rightarrow m\pm 1}
    =
    [S(S+1)-m(m\pm 1)] \Gamma_\pm
\end{equation}
with
\begin{equation}\label{eq:GammaPM}
   \Gamma_\pm
   =
   \frac{1}{4\hbar} \int d\omega
   \zeta_\mp(-\omega) \check{C}_\mp(\omega)
  \,.
\end{equation}
All other rates vanish. Note, that the definition of $\Gamma_\pm$
in this work is different from that employed in
Ref.~\cite{GeorgErelaxPRB02}. The rate equations for the occupation
probabilities
\begin{eqnarray}\label{eq:RateEq}
 \dot{\cP}^{}_m
 &=&
 -\Gamma_{m \rightarrow m + 1} \cP^{}_m
 -\Gamma_{m \rightarrow m - 1} \cP^{}_m
\nonumber \\
&&
 +\Gamma_{m+1 \rightarrow m} \cP^{}_{m+1}
 +\Gamma_{m-1 \rightarrow m} \cP^{}_{m-1}
\end{eqnarray}
with the normalization condition $\sum_m \cP^{}_m \equiv 1$
form a closed set of equations with the steady 
state solution
\begin{equation}\label{eq:OccProb}
    \cP^{}_m
    =
    \frac{\Gamma_+^{M/2+m} \Gamma_-^{M/2-m}}
    {\sum_{n=0}^M \Gamma_+^{M-n} \Gamma_-^{n}}
 \, .
\end{equation}
The probabilities obey the
obvious balance relation $\cP^{}_m/\cP^{}_{m+1}=\Gamma_-/\Gamma_+$
which leads to the thermal distribution
in equilibrium.

At vanishing magnetic field, $B=0$, the probabilities are all
equal, $\cP_m=1/(M+1)$, and the equal time correlation functions
read $C_z(t=0)=S(S+1)/3$ and $C_\pm(t=0)=2 C_z(t=0)$. If in
addition the distribution functions are spin independent, the
renormalized coupling constants become process independent
$J^z_\pm = J^\pm \equiv J(\epsilon)$, and the auxiliary functions
read $\zeta_z=2\zeta_\pm\equiv 2\zeta(\omega)$. Inserting this 
in the correlation functions, we find 
$C(\omega)=[C_+(\omega) + C_-(\omega)]/2+C_z(\omega)=3C_+(\omega)/2$. 
In equilibrium and at low temperatures the
width shrinks to zero and leads to $C(\omega)\rightarrow 2\pi
S(S+1) \delta(\omega)$.

The inelastic relaxation rate $1/\tau_{\text{inel}}$ at $B=0$ 
is the spin--flip rate $1/\tau_{\text{sf}}$ reduced by the 
quasi--elastic rate, and in general we have
$1/\tau_{\text{inel}} < 1/\tau_{\text{sf}}$. 
Quite generally, due to a sum rule for the spin--spin
correlation function, the spin--flip rate obeys
\begin{equation}
 \frac{1}{\tau_{\text{sf}}}
=
 \frac{1}{2} \sum_{\sigma, \sigma'} \int d\epsilon \,
 W_{\sigma, \sigma'}(\epsilon,\epsilon')
=
 \frac{\pi}{2 \hbar} \frac{c_{\text{imp}}}{\rho}
 \left( \rho J \right)^2    S(S+1) \, .
\label{eq:spinfliprate}
\end{equation}
In order to discuss
the possibility of reducing the impurity density by increasing the
spin $S$ at constant inelastic electronic rate we may as well 
consider the spin--flip rate. As already explained 
in the introduction, Eq.\ $(\ref{eq:spinfliprate})$ suggests 
a decrease of the impurity
density with increasing spin $S$. This is true only if the
renormalization of the coupling constants is independent of $S$
meaning at temperatures much higher than the Kondo temperature. To
explain the experiments, however, $\rho J$ has to be around $1/3$
to be almost voltage independent. Otherwise the renormalization 
wouldn't allow for the experimentally observed scaling property 
of the distribution function 
$f(\epsilon, e U)= f(\epsilon/eU)$, see
Ref.~\cite{GeorgErelaxPRB01,GeorgErelaxPRB02}. In this regime,
however, the renormalization depends on the spin $S$ and scales
for large spin like $\rho J \sim 1/\sqrt{\pi^2 S(S+1)}$ leading to
a spin independent rate $1/\tau_{\text{sf}}$. Actually, the
renormalized coupling constant equals the spin--flip t--matrix
\cite{GeorgErelaxPRB01} which obeys a unitarity condition. It
reaches a maximum, $\rho J = 1/\sqrt{\pi^2 S(S+1)}$, at the
Kondo temperature where the rate again would become independent 
of spin for all $S$. Although our theory is no longer valid in 
this regime, the outcome is quite physical since electrons 
always transfer the same spin when scattering from one impurity
independent of $S$. This shows that using our theory
an increase of the spin does not lower the impurity 
concentrations needed to describe the experiments. Even a more 
involved theory valid below Kondo temperature is not likely 
to help much since the
scattering rate cannot exceed the limit discussed above.

To discuss the magnetic field dependence of the rates $W$ we
consider two limiting cases. For low magnetic fields where the
Zeeman splitting $E_H$ is much smaller than the temperature or the
applied voltage the occupation probabilities are all of the same
order. Also the lifetimes do not change much and the behavior is
dominated by the shift in the spin--flip correlation functions,
$\omega \rightarrow \omega \pm E_H$. Therefore, there is no
dependence on the spin $S$ for small magnetic fields. For higher
magnetic fields of the order of temperature or applied voltage,
higher spin states are rapidly depopulated so that only two spin
states like in the $S=1/2$ case lead to the dominant contribution.
For higher $S$ this is of course just a fraction and therefore in
this regime the rates are even smaller than in the $S=1/2$ case.

In this work we have studied electron relaxation rates caused 
by magnetic
impurities of arbitrary spin generalizing recent results for
$S=1/2$. It is found that an increase of the spin $S$ does not
change the qualitative outcome and the rate at vanishing magnetic
field is even unaffected by the spin for large $S$. Therefore,
assuming magnetic impurities with higher spin $S$ does not resolve 
the disagreement between Kondo impurity densities determined 
by energy relaxation experiments
and weak localization experiments.
The authors would like to thank B.~L.~Altshuler, A.~Anthore,
Y.~M.~Galperin, F.~Pierre, and H.~Pothier
for valuable discussions.
Financial support was provided by the Deutsche
Forschungsgemeinschaft (DFG).
\vspace{-0.3truecm}

%
%

\end{document}